\documentclass[prd,twocolumn,superscriptaddress,altaffilletter,amssymb,showpacs,nofootinbib]{revtex4}
\usepackage[dvips]{graphicx}
\usepackage{amsmath}
\newcommand{\be}{\begin{equation}}
\newcommand{\ee}{\end{equation}}
\newcommand{\bea}{\begin{eqnarray}}
\newcommand{\eea}{\end{eqnarray}}
\newcommand{\vphi}{\varphi}

\begin{document}

\title{A Note on de Sitter Embedding of $f(R)$ Theories}

\author{Israel Quiros}\email{israel@uclv.edu.cu}
\affiliation{Departamento de F\'{\i}sica, Universidad Central de
Las Villas, 54830 Santa Clara, Cuba.}

\author{Yoelsy Leyva}\email{yoelsy@uclv.edu.cu}
\affiliation{Departamento de F\'{\i}sica, Universidad Central de
Las Villas, 54830 Santa Clara, Cuba.}

\author{Yunelsy Napoles}\email{yna@uclv.edu.cu}
\affiliation{Departamento de Matem\'{a}tica, Universidad Central de
Las Villas, 54830 Santa Clara, Cuba.}

\date{\today}
\begin{abstract}
The consequences of the constraints which de Sitter embedding of $f(R)$ theories imposes on the Lagrangian's parameters, are investigated within the metric formalism. It is shown, in particular, that several common $f(R)$ Lagrangians do not actually admit such an embedding. Otherwise, asymptotic matching of local solutions of the corresponding models with background (maximally symmetric) spaces of constant curvature is either unstable or, anti-de Sitter embedding is the only stable embedding. Additional arguments are given in favour of a previous claim that a class of $f(R)$ models comprising both positive and negative powers of $R$ (two different mass scales), could be a nice scenario where to address, in a united picture, both early-time inflation and late-time accelerated expansion of the universe. The approach undertaken here is used, also, to check ghost-freedom of a Dirac-Born-Infeld modification of general relativity previously studied in the bibliography. 
\end{abstract}


\maketitle

\section{Introduction}

Attempts to modify the Einstein-Hilbert (EH) action of general relativity (GR) $$S_{EH}=\frac{1}{2\kappa^2}\int d^4x\sqrt{|g|}\left(R-2\Lambda\right),$$ where $R=g^{\mu\nu}R_{\mu\nu}$ is the Ricci curvature scalar, and $\Lambda$- the cosmological constant ($\kappa^2=m_{Pl}^{-2}=8\pi G$), have been motivated by a number of reasons. In particular, renormalization at one-loop demands that the Einstein-Hilbert action be supplemented by higher order curvature terms \cite{udw}.\footnote{Higher order actions are indeed renormalizable (but not unitary) \cite{stelle}).} Besides, when quantum corrections or string theory are taken into account, the effective low energy action for pure gravity admits higher order curvature invariants \cite{qstring}. 

More recently it has been suggested that the present cosmic speedup could have its origin in -- among other possibilities -- corrections to the GR equations of motion, generated by non-linear contributions of the scalar curvature $R$ in the pure gravity Lagrangian of $f(R)$ theories \cite{odintsov1,frspeedup,positive,carroll}. It has been demonstrated, however, that Solar system experiments seem to rule out $f(R)$ theories that are able to accommodate present accelerated expansion of the Universe \cite{olmo2005,olmoprd2007} (for recent reviews see \cite{odintsovrev,faraonirev,faraoni2008,sotiriou2008}). The demonstration relies on the weak-field limit expansion of the $f(R)$ Lagrangian, and the consequent calculation of post-Newtonian contributions to the metric coefficients \cite{olmoprd2007,faraonirev,olmo2005}.

Nonetheless, even if $f(R)$ theories are not a viable alternative to explain current acceleration of the expansion, their relevance to study early-time inflation \cite{starobinsky} might fuel further interest in these alternatives to general relativity.

In the present note we aim at investigating the embedding of $f(R)$ theories in the ambient background space-time, an issue that is central for the study of the weak-field limit and of the post-Newtonian metric of these theories. Our study will rely on the metric variational formalism. We shall focus, in particular, on the constraints de Sitter embedding of $f(R)$ theories \cite{odintsovdesitter} imposes on the Lagrangian's overall parameters. Although the conditions for the embedding are usually assumed to be obeyed, thorough consideration of these conditions in standard calculations of the relevant parameters is mostly lacking. Thorough consideration of the condition that makes a de Sitter embedding possible, may save additional (mostly unnecessary) assumptions on the cosmic dynamics to reach to robust physical conclusions.

In order to judge about the theoretical viability of a given $f(R)$ gravity model, the constraints imposed by matching of the local solutions of the equations of the theory with asymptotic solutions of constant curvature $R=R_c$ -- in a maximally symmetric background --, are used to fix the effective parameters that characterize the non-linearities of $f(R)$ theories, such as the effective gravitational coupling and the effective mass of the additional scalar degree of freedom $\phi=f'(R)$.

It will be shown, in particular, that several common $f(R)$ Lagrangians either do not actually admit a stable de Sitter (dS) embedding, or admit an stable asymptotic matching of local solutions with the ambient metric of an anti-de Sitter (AdS) space.

\section{$f(R)$ Equations of Motion}

In this note we shall focus in gravity theories that can be derived from the general action

\be S_{tot}=S_g+S_{mat}(g_{\mu\nu},\chi),\label{action}\ee where $S_{mat}$ is the action of the matter degrees of freedom -- collectivelly denoted by $\chi$ --, while the pure gravity action $S_g$ has the form of an $f(R)$ theory \cite{faraonirev}:

\be S_g=\frac{1}{2\kappa^2}\int d^4x \sqrt{|g|}f(R).\label{fr action}\ee 

The field equations that can be derived from the action (\ref{action}) through the standard metric variational procedure, are the following:

\be f' R_{\mu\nu}-\frac{1}{2}f g_{\mu\nu}-\left(\nabla_\mu\nabla_\nu-g_{\mu\nu}\Box\right)f'=\kappa^2 T_{\mu\nu},\label{feqs}\ee where $f'\equiv df/dR$, and, as customary, the matter stress-energy tensor $T_{\mu\nu}$ is defined through: $$T_{\mu\nu}=-\frac{2}{\sqrt{|g|}}\frac{\delta S_{mat}}{\delta g^{\mu\nu}}.$$ The trace of equation (\ref{feqs}) yields to a dynamical equation for determining of the curvature scalar:

\be f' R-2f+3\Box f'=\kappa^2 T.\label{trace}\ee 

For purposes of comparison with canonical (EH) general relativity, equations (\ref{feqs}) can be re-arranged in the following way:

\bea &&R_{\mu\nu}-\frac{1}{2}g_{\mu\nu}R=8\pi G_{eff}\left(T_{\mu\nu}+T_{\mu\nu}^{eff}\right),\label{grfeqs}\\
&&\kappa^2 T_{\mu\nu}^{eff}=\left(\nabla_\mu\nabla_\nu-g_{\mu\nu}\Box\right)f'+\frac{1}{2}\left(f-f'R\right)g_{\mu\nu},\nonumber\eea where $8\pi G_{eff}=\kappa^2/f'$ is the effective gravitational coupling. The right-hand-side (RHS) of (\ref{grfeqs}) can now be seen as the source terms for the metric, meaning that the metric is generated by the matter and by non-linear terms related to the scalar curvature \cite{olmoprd2007}. Additionally, the scalar curvature $R$ satisfies a second-order differential equation (Eq.(\ref{trace})) with the trace $T=g^{\mu\nu}T_{\mu\nu}$ of the matter stress-energy tensor and other (non-linear) curvature terms acting as sources. In other words, the Ricci scalar curvature is now a dynamical quantity whose dynamics is determined by the trace equation (\ref{trace}).

\section{Boundary Conditions}

Here we shall focus on the embedding of a local (non-compact) object in the ambient background space-time, within the metric approach to $f(R)$ theories. This issue is motivated by the fact that, in order to have a complete description of the local physical system, one has to take into account, also, its interaction with the environment -- in the present case the rest of the universe \cite{olmoprd2007}.

The boundary conditions for the metric are imposed by a suitable choice of coordinates. For instance, one might consider that asymptotically (far away from the local source) the metric is Minkowski and fix its first derivatives to zero \cite{will}. Since, according to the field equations (\ref{feqs}), (\ref{trace}) $f'(R)$ is now a dinamical quantity, one has to impose additional boundary conditions on $f'$ also. In other words, the local system will interact with the background cosmology via the boundary value $f'(R_c)$ ($R_c$ is the cosmic value of the Ricci curvature scalar) and its time derivative. To fix the latter boundary condition one may invoke the adiabatic approximation, according to which the evolution of the universe is very slow when compared with the local dynamics, so that we can ignore terms such as $\dot f'(R_c)$ (the dot means derivative with respect to the cosmic time).

In order to derive solutions of the equations (\ref{feqs}) (or, alternativelly, of Eq.(\ref{grfeqs})), one has to expand the metric about the asymptotic Minkowski metric $\eta_{\mu\nu}$: $$g_{\mu\nu}=\eta_{\mu\nu}+h_{\mu\nu}(t,x),$$ where $h_{\mu\nu}$ represent small perturbations about the background ($|h_{\mu\nu}|\ll 1$), and, at the same time, we expand the scalar field degree of freedom $\phi=f'$: $\phi=\phi_c+\varphi(t,x)$. The trace equation tends to $$3\Box_c f'_c+f'_c R_c-2f_c=\kappa^2 T_c,$$ where, for a vacuum background $T_c=0$. Here $f_c=f(R_c)$, $f'_c=f'(R_c)$, etc. One may calculate the effective mass squared of the scalar field perturbation propagating in the background \cite{olmo2005} (see also \cite{additional}):\footnote{For simple models the expression for determining of the effective mass squared was studied in \cite{odintsov}.}

\be m_\varphi^2=\frac{f'_c-R_c f''_c}{3f''_c},\label{mass}\ee which will be used here as a criterion to judge the physical viability of the theory. Only scalar field perturbations with positive $m_\varphi^2>0$ are physical (the perturbations exponentially damp). Otherwise, for negative $m_\varphi^2<0$, the perturbation of the scalar field degree of freedom oscillate. It has been demonstrated that the corresponding solutions are unphysical \cite{olmo2005}.

\section{Stability Issue}

An independent criterion to evaluate the viability of a given embedding is based on Ricci stability of the $f(R)$ theories. The stability condition $f''(R_c)\geq 0$ expresses the fact that the scalar degree of freedom $\phi(R)=f'(R)$ is not a ghost \cite{faraoni2007}. Additionally, studies of linear stability of the scalar perturbation lead to the following condition \cite{faraonirev}: 

\be f'_c\geq R_c f''_c.\label{linear stability}\ee

Simultaneous requirement of linear and of Ricci stability $$f'_c\geq R_c f''_c,\;\;f''_c\geq 0,$$ lead to positivity of the effective mass squared of the scalar field degree of freedom ($m_\vphi^2\geq 0$). Notice, however, that the contrary statement is not true in general. Actually, the mass squared is positive also if, simultaneously, $f'_c< R_c f''_c$, and $f''_c< 0$.

In the present note, to avoid misleading results, we will use as criteria to judge about the stability of the de Sitter embedding the following requirements: i) positivity of the mass squared $m_\vphi^2\geq 0$, and ii) Ricci stability $f''_c\geq 0$. Joint fulfillment of the above requirements means that linear stability is also granted.

\section{de Sitter Embedding}

Our goal will be to fix the effective parameters that characterize the non-linearities of $f(R)$ theories (e.g., the effective gravitational coupling and the effective mass $m_\vphi^2$), around solutions with constant curvature $R=R_c$ in a maximally symmetric background \cite{wands}:

\be R_{\mu\nu\sigma\upsilon}=\frac{R_c}{12}\left(g_{\mu\sigma} g_{\nu\upsilon}-g_{\mu\upsilon} g_{\nu\sigma}\right)\;\Rightarrow\;R_{\mu\nu}=\frac{R_c}{4}g_{\mu\nu}.\label{desitter}\ee 

According to the trace equation (\ref{trace}) the condition for the existence of such a embedding can be written in the following way \cite{faraonirev}:\footnote{The importance of condition (\ref{condition}) to exit from matter dominance with positive conclusion about following acceleration in many $f(R)$ theories was studied in \cite{troisi}. This result contradicts the one in \cite{amendola} stating that it is impossible in general.}

\be f'_c R_c-2f_c=0,\;\Rightarrow\;R_c=2\frac{f_c}{f'_c}.\label{condition}\ee 

The above condition, together with the requirement of positivity of the mass squared $m_\vphi^2\geq 0$, and of Ricci stability $f''_c\geq 0$, are central in the subsequent discussion since, in general, Eq.(\ref{condition}) reduces to an algebraic constraint on the overall parameters of the $f(R)$ Lagrangian. 

Actually, take, for instance, $f(R)$ theories that can be written in the form $f(R)=R+\alpha g(R)$, where $\alpha$ is a small parameter.\footnote{GR is obtained as the limit $\alpha\rightarrow 0$ of these theories.} The condition (\ref{condition}) translates into the following constraint on the boundary value of $g'$:

\be g'_c-2\frac{g_c}{R_c}=\frac{1}{\alpha}.\label{alpha}\ee 

To see how the above condition constraints the overall parameters of the given theory, as an illustration, let us consider the theory $f(R)=R+\alpha R^2$ \cite{starobinsky}. In this case $g(R)=R^2,\;\Rightarrow\;g'_c=2R_c$, while $g_c/R_c=R_c$, so that the condition (\ref{alpha}) implies that $$g'_c-2\frac{g_c}{R_c}=0=\frac{1}{\alpha}\;\Rightarrow\;\alpha=\infty,$$ contrary to the requirement that $\alpha$ be a small quantity. This means that matching of local solutions with asymptotic vacuum solutions of constant curvature, in the model of Ref.\cite{starobinsky}, is not allowed.  

A somewhat similar conclusion is attained for generic models with $g(R)=R^n$ ($n>2$). In this case the condition for the embedding of local solutions into background spaces of constant curvature, leads to: $$R_c=\left[\frac{1}{(n-2)\alpha}\right]^{1/n-1},$$ so that, since in general $\alpha$ is a very small mass scale, the embedding can be consistent only at high curvature.

\section{Generic $f(R)$ Theories}

As additional illustrations of the importance of the condition (\ref{condition}) to judge about the viability of de Sitter embeddings, in this section we shall explore several classes of generic $f(R)$ theories that has been extensivelly studied, for instance, in references \cite{positive,olmo2005,amendola,faraonirev,negative}.

\subsection{Positive Powers of $R$}

Let us consider theories given by \cite{positive,olmo2005}:

\be f(R)=R+\frac{\epsilon R^n}{M^{2n-2}},\label{positive powers}\ee where $M$ represents a very large mass scale, $\epsilon\equiv\pm 1\;\Rightarrow\;\epsilon^2=1$, and $n\geq 0$. The function $f(R)$ in Eq.(\ref{positive powers}) comprises several cases formerly studied, for instance, in Ref.\cite{starobinsky,amendola}. The model of \cite{starobinsky}, for instance, is recovered if in (\ref{positive powers}) one sets $n=2$.

In this case the constraint (\ref{condition}) can be written as: $$2\frac{f_c}{f'_c}=\frac{2M^{2n-2}R_c+2\epsilon R_c^n}{M^{2n-2}+\epsilon nR^{n-1}}=R_c,$$ which leads to the following relationship of the boundary value $R_c$ with the parameter $M$:
 
\be \left(\frac{M^2}{R_c}\right)^{n-1}=\epsilon(n-2).\label{rc/m2}\ee On the other hand, for the effective mass squared of the scalar field perturbation (\ref{mass}), one obtains \cite{olmo2005}:

\be m_\varphi^2=\frac{R_c}{3\epsilon(n-1)}\left[\frac{1}{n}\left(\frac{M^2}{R_c}\right)^{n-1}-\epsilon(n-2)\right].\label{m2}\ee By susbtituting (\ref{rc/m2}) in (\ref{m2}), one is led to the following expression for the mass squared:

\be m_\varphi^2=-\left(\frac{n-2}{3n}\right)R_c,\ee while, on the other hand

\be f''_c=\frac{\epsilon n(n-1)}{R_c}\left(\frac{R_c}{M^2}\right)^{n-1}=\frac{n}{R_c}\left(\frac{n-1}{n-2}\right).\ee 

Notice that in the above expressions for $m_\vphi^2$ and $f''_c$ there is no explicit dependence on $\epsilon$. Otherwise, the results of the stability study are independent of the sign of the second term in the right-hand-side (RHS) of equation (\ref{positive powers}).

As clearly seen, for $n>2$, a de Sitter embedding of local solutions ($R_c>0$) is unstable since the requirements of positivity of mass squared and of Ricci stability can not be simultaneously met. The same is true for an anti-de Sitter embedding ($R_c<0$). To state it more clearly, for the region of parameter space $n>2$, $f(R)$ models given by (\ref{positive powers}) do not admit matching of local solutions with asymptotic (vacuum, maximally symmetric) spaces of constant curvature. Worth noticing that the same conclusion holds true for $1<n<2$ since, in this case, $m_\vphi^2$ and $f''_c$ are of oposite sign.

The above matching is possible only for the region of parameter space $0\leq n< 1$. This is an important result since, as discussed in \cite{faraonirev}, theories with $f(R)$ given by (\ref{positive powers}) are compatible with the observations, precisely, in the region of the parameter space $0<n\leq 0.25$. In this case, however, $M$ has to be a sufficiently small mass scale.

\subsection{Negative Powers of $R$}

An interesting alternative to (\ref{positive powers}) can be given by \cite{olmo2005}:

\be f(R)=R+\frac{\epsilon\mu^{2n+2}}{R^n},\label{negative powers}\ee where, as before $\epsilon=\pm 1$, and $\mu^2$ is a tiny mass scale ($n\geq 0$). The above expression for $f(R)$ contains as a particular case, for instance, the one studied in \cite{negative}. In this case the constraint (\ref{condition}) translates into:

\be 2\frac{f_c}{f'_c}=\frac{2R_c+2\epsilon\mu^{2n+2}/R_c^n}{1-\epsilon\mu^{2n+2}/R_c^{n+1}}=R_c.\ee This, in turn, yields to the following relationship:

\be \left(\frac{R_c}{\mu^2}\right)^{n+1}=-\epsilon(n+2),\ee which, in general, holds true for the negative sign $\epsilon=-1$. When the above constraint is substituted in the expression for the effective mass squared of the scalar perturbation \cite{olmo2005},

\be m_\varphi^2=\frac{R_c}{3\epsilon(n+1)}\left[\frac{1}{n}\left(\frac{R_c}{\mu^2}\right)^{n+1}-\epsilon(n+2)\right],\ee one gets the following expression: 

\be m_\varphi^2=-\left(\frac{n+2}{3n}\right)R_c,\label{m'}\ee while, on the other hand,

\be f''_c=-\left(\frac{n+1}{n+2}\right)\frac{n}{R_c}.\label{f''}\ee 

Notice, as in the former case, that the expressions for $m_\vphi^2$ and $f''_c$ do not contain $\epsilon$, meaning that stability of the de Sitter embedding does not depend on the sign of the second term in the RHS of Eq.(\ref{negative powers}).

From equations (\ref{m'}), (\ref{f''}) -- considering the requirements $m_\vphi^2\geq 0,\;f_c''\geq 0$ -- it is evident that a dS embedding is unstable, while an AdS embedding of local solutions is stable instead. In consequence, local solutions of the equations of $f(R)$ theory given by (\ref{negative powers}), can be consistently matched only with asymptotic AdS background space. 

The latter result represents an additional argument in favour of previous claims that $f(R)$ theories do not seem a good alternative to explain the late-time cosmic speed-up \cite{olmo2005}(see the related discussion in section VII).

\subsection{Combined Powers of $R$}

A natural generalization of models given by (\ref{positive powers}) and (\ref{negative powers}), can be based on the following form of the $f(R)$ function \cite{odintsov}:

\be f(R)=R\pm\frac{\mu^{2n+2}}{R^n}+\frac{R^m}{M^{2m-2}},\label{combined powers}\ee where, as before, $\mu$ and $M$ are a small and a large mass scales respectively, and we shall consider that $n\geq 0$, $m\geq 2$. The above expression contains as a particular case the function \cite{faraonirev}: $$f(R)=R-\frac{\mu^4}{R}+\alpha R^2.$$

The condition for embedding of local solutions in an asymptotic space of constant curvature (\ref{condition}) can be written, in the present case, in the following form:

\be \left(\frac{R_c}{M^2}\right)^{m-1}=\frac{1\pm(n+2)\left(\mu^2/R_c\right)^{n+1}}{m-2}.\label{condition'}\ee On the other hand, the effective mass squared of the scalar perturbation is given by: $$ m_\vphi^2=\frac{1\pm n(n+2)\left(\frac{\mu^2}{R_c}\right)^{n+1}+m(m-2)\left(\frac{R_c}{M^2}\right)^{m-1}}{3f''(R_c)},$$ where $$f''(R_c)=\frac{\pm n(n+1)}{R_c}\left(\frac{\mu^2}{R_c}\right)^{n+1}+\frac{m(m-1)}{R_c}\left(\frac{R_c}{M^2}\right)^{m-1}.$$ When the condition (\ref{condition'}) is substituted in the above expressions one gets:

\be m_\vphi^2=\frac{m+1\pm(m-n)(n+2)\left(\mu^2/R_c\right)^{n+1}}{3f''(R_c)},\ee
\be f''(R_c)=\frac{m(m-1)}{R_c(m-2)}\left[1\pm k \left(\frac{\mu^2}{R_c}\right)^{n+1}\right],\ee where, for shorten, we have introduced a constant parameter: $$k\equiv\frac{n(n+1)(m-2)+m(m-1)(n+2)}{m(m-1)},$$ and we shall consider only situations where $m>n$. 

It is evident from the above equations that, when the plus sign in the second term of the RHS of Eq. (\ref{combined powers}) is considered, a de Sitter embedding of local solutions of the $f(R)$ theory is stable. 

If one chooses, instead, the negative sign in (\ref{combined powers}): $$f(R)=R-\frac{\mu^{2n+2}}{R^n}+\frac{R^m}{M^{2m-2}},$$ then, since as it can be demonstrated: $$k>\frac{(m-n)(n+2)}{m+1},$$ it follows that stability of a de Sitter embedding of local solutions of the corresponding $f(R)$ theory can be achieved only if $(R_c/\mu^2)^{n+1}\geq k$. 

Notice that, in the above discussion, the case with $m=2$ can not be considered due to indefinition of several expressions. Notwithstanding, this case (with $n=1$) has been studied in Ref.\cite{faraonirev}.

\section{Dirac-Born-Infeld Modification of General Relativity}

To be phenomenologically viable, non-linear modifications of general relativity have to satisfy several physically motivated requirements \cite{deser}: 

\begin{enumerate}

\item Reduction to EH action at small curvature,

\item ghost freedom,

\item regularization of some singularities (as, for instance, the Coulomb-like Schwarzschild singularity), and

\item supersymmetrizability.

\end{enumerate} The later requirement is quite stringent and, for most purposes, might be excluded. A theory that fulfills the above requirements can be based on the following Dirac-Born-Infeld-type action \cite{comelli}:

\be S=\frac{1}{\kappa^4}\int d^4x\sqrt{|g|}\left(1-\sqrt{1-\alpha\kappa^2 R+\beta\kappa^4 {\cal E}}\right),\label{action comelli}\ee where ${\cal E}\equiv R^2-4R_{\mu\nu}R^{\mu\nu}+R_{\mu\nu\sigma\upsilon}R^{{\mu\nu\sigma\upsilon}}$ is the Gauss-Bonnet term. It has been demonstrated in Ref.\cite{comelli} that this action has the EH leading term at small curvature, is ghost-free and, for an appropiate region in the parameter space, it shows indications for the cancellation of the Coulomb-like Schwarzschild singularity. 

In this section, for simplicity, we shall focus in the following Dirac-Born-Infeld (DBI) modification of the Einstein-Hilbert action:

\be S_g=\frac{1}{\kappa^4}\int d^4x \sqrt{|g|}\left(1-\sqrt{1-\alpha\kappa^2 R}\right).\label{dbi action}\ee Notice that this modification of the Einstein-Hilbert action is a particular case of (\ref{action comelli}) for $\beta=0$. In consequence (\ref{dbi action}) fulfills the requirements $1-4$ above. In particular, as it was demonstrated in Ref.\cite{comelli}, it has the correct EH limit at low curvature and is free of ghosts.

Through using the approach of \cite{olmo2005,olmoprd2007} (see also \cite{faraonirev}), we will show here that the above statement about ghost freedom is wrong for positive values of the parameter $\alpha>0$. Actually, the theory given by (\ref{dbi action}) can be recast, alternatively, into the form of an $f(R)$ theory of the kind in Eq.(\ref{fr action}), with 

\be f(R)=\frac{2}{\kappa^2}\left(1-\sqrt{1-\alpha\kappa^2 R}\right).\label{dbi-fr}\ee The condition (\ref{condition}) for the embedding of local solutions of the above $f(R)$ theory into a maximally symmetric (vacuum) background space of constant curvature, leads to the following relationship between the boundary value of the curvature $R_c$ and the parameters $\alpha$, $\kappa^2$:

\be 2\frac{f_c}{f'_c}=R_c\;\;\Rightarrow\;\;R_c=\frac{8}{9}\frac{1}{\alpha\kappa^2}.\label{condition''}\ee When this boundary $R$-value is substituted into the definition of effective mass squared of the scalar field degree of freedom:

\be m_\vphi^2=\frac{f'_c-R_c f''_c}{3 f''_c}=-\frac{R_c}{4}\;\left(=-\frac{2}{9}\frac{1}{\alpha\kappa^2}\right),\label{mass'}\ee where we have taken into account that $f''_c=27\alpha^2\kappa^2/2$ -- the theory is Ricci stable for any value of the overall parameters, it can be seen from (\ref{mass'}) that the theory (\ref{dbi action}) can be consistently matched only with AdS background ($\alpha<0$). Otherwise, for positive $\alpha>0$, the scalar degree of freedom carrying the non-linear $R$-contribution, is a (unphysical) ghost degree of freedom.

\section{Discussion}

An accurate handling of boundary conditions in $f(R)$ theories requires to solve equations (\ref{feqs}) (or (\ref{grfeqs})), and (\ref{trace}), for the metric and the scalar field $\phi=f'(R)$ respectively, in the cosmic regime, where homogeneity and isotropy lead to a Friedman-Robertson-Walker metric $g^c_{\mu\nu}=\text{diag}(-1,a(t)^2\delta_{ij})$, and to a space-independent value of the scalar field $\phi_c=\phi_c(t)$ \cite{olmo2005}. At smaller scales local deviations from the cosmic values of the fields become appreciable. Then, usually, one invokes the weak-field limit and treats these deviations as small perturbations around the background boundary values $g^c_{\mu\nu}$, $\phi_c$ \cite{olmo2005,olmoprd2007,will}. 

Hence, it is clear the role stability of de Sitter embeddings plays in the phenomenology of the $f(R)$ models: they act as a selection rule to choose a theory whose local solutions can be matched with phenomenologically viable cosmological models. This is the reason why we chose to look for stable matching of local solutions with asymptotic -- maximally symmetric -- background spaces of constant (positive) curvature $R_c>0$: In a cosmological setting such spaces fit well the existing observational evidence on late-time accelerated pace of the cosmic expansion. 

In the present note we investigate the constraints conditions for an appropiated embedding in a de Sitter (vacuum) background, impose on the parameters of the $f(R)$ Lagrangian. Although the conditions for the embedding are usually assumed to be obeyed, thorough consideration of these conditions in standard calculations of the relevant parameters is mostly lacking. In Ref.\cite{olmo2005}, for instance, the expressions for the effective mass squared of the scalar field degree of freedom are given, but the condition for consistency of the embedding are not substituted into these expressions so that, in consequence, the author was able to make physical conclusions only after assuming a given cosmological solution (for matter-dominated era, in particular). On the contrary, thorough consideration of the condition (\ref{condition}) for the embedding may save additional (mostly unnecessary) assumptions on the cosmic dynamics, to reach to robust physical conclusions.

$f(R)$ models were primarily intended to explain late-time cosmic acceleration. The reasoning behind this effect is that, typically, during the course of the cosmic expansion, the curvature dilutes and the term non-linear in $R$ starts dominating the late-time cosmic dynamics, acting as a dark energy source encoded in the effective stress-energy tensor $T_{\mu\nu}^{eff}$ in the RHS of Eq.(\ref{grfeqs}). However, there are several $f(R)$ models that fail to be compatible with late-time dynamics. The approach undertaken by us in this note allows one to have an additional check of consistency. Take, for instance, $f(R)$ models given by (\ref{negative powers}): $$f(R)=R+\epsilon\frac{\mu^{2n+2}}{R^n},$$ where $\epsilon=\pm 1$. The above models have been shown in the former section (VI.B) to have a stable anti-de Sitter embedding. Otherwise, the corresponding theories can not be embedded in a de Sitter space, which means, in turn, that any solution of the equations of the theory that is compatible with cosmic acceleration has to be, neccessarily, unstable. This result may be considered as an additional criterion to judge about the phenomenological viability of $f(R)$ theories given by (\ref{negative powers}).

For models where the function $f(R)$ is given by (\ref{positive powers}): $$f(R)=R+\epsilon\frac{R^n}{M^{2n-2}},$$ one may found a region in parameter space $0\leq n< 1$, where a de Sitter embedding is stable. As long as we know, no such conclusive argument has been given before on $f(R)$ theories of the kind (\ref{positive powers}). A lucky circunstance is related with the fact that, the region of parameter space where the model of (\ref{positive powers}) is compatible with the cosmological observations $0<n\leq 0.25$, is contained within the region that allows stable de Sitter embedding. 

In both cases -- $f(R)$ theories given by either (\ref{positive powers}), or (\ref{negative powers}) --, the stability of a de Sitter embedding is independent on the sign $\epsilon$ of the non-linear $R$-term. Also a robust conclusion on the parameter space of the above theories.

There are other $f(R)$ models as, for instance, the one of Ref.\cite{starobinsky}: $$f(R)=R+\alpha R^2,$$ which do not admit embedding of local solutions in a maximally symmetric background space of constant curvature.\footnote{Besides leading to early-time inflation and satisfying the solar system observational constraints, the $R^2$-inflation model of Ref.\cite{starobinsky}, also seems compatible with CMBR observations \cite{cmbr}.}

Additional comments deserve the $f(R)$ models given by Eq.(\ref{combined powers}) \cite{odintsov}, which contain combined powers of $R$. As properly noticed, as long as one considers only the positive sign in front of the second term in the RHS of (\ref{combined powers}): $$f(R)=R+\frac{\mu^{2n+2}}{R^n}+\frac{R^m}{M^{2m-2}},$$ the corresponding $f(R)$ models admit a stable de Sitter embedding, so that these could be appropriated to address late-time accelerated expansion. Due to the existence of two mass scales: a large one fixed by $M$, which sets the scale at which early-time inflation happens, and a tiny one fixed by $\mu$, which sets the (inverse of the) length scale at which late-time accelerated expansion occurs, the above models could represent interesting alternatives to address unified description of early-time inflation (without the inflaton), and late-time accelerated expansion of the universe (without dark energy). At high enough energy the third term $\propto R^m$ in the RHS of the above equation causes the universe to inflate, while the second term $\propto R^{-n}$ dominates at late times. 

The approach undertaken in this note, permitted us to demonstrate also, that the DBI modification of general relativity proposed in \cite{comelli} (in fact a particular case of it given by the action (\ref{dbi action})), is free of ghosts only for negative values of the parameter $\alpha<0$. The possible conflict of this result with the result of Ref.\cite{comelli} can be based on the fact that, in order to garrant ghost-freedom, the author of \cite{comelli} only demanded that the mass of the additional spin-two degree of freedom (not present in our simpler case) $m_2\rightarrow\infty$. No additional criterion was given on the mass of the spin-zero degree of freedom $m_0^2$ ($m_\vphi^2$ in this note), which is central in the present discussion.

\section{Conclusions}

In this note we have explored the phenomenological consequences of the thorough consideration of the conditions for a stable de Sitter embedding of the local solutions of a wide class of $f(R)$ theories. Otherwise, the interaction of the local system with the background cosmology, imposes boundary conditions which, when taken into accout, impose constraints on the overall parameters of the $f(R)$ Lagrangian. Although it is a well-known fact, it is not thoroughly used to constraint the parameter space of the models.

It has been shown here that a wide variety of models do not actually pass the de Sitter stability test. There are models which admit anti-de Sitter embedding of local solutions in the background cosmology. Other, in general, do not admit stable matching of local solutions with an asymptotic, maximally symmetric background of constant curvature $R_{\mu\nu}=R_c g_{\mu\nu}/4$. Additionally, it has been shown here that a DBI modification of general relativity formerly studied in Ref.\cite{comelli} is ghost-free only for negative values of one of the parameters of the Lagrangian ($\alpha<0$). This result conflicts with the one in \cite{comelli}, claiming that theories based on Lagrangians of the kind (\ref{action comelli}) -- which include as a particular case the one given by (\ref{dbi action}) -- are ghost-free in general.

There is a class of $f(R)$ models, where non-linear $R$-terms have positive, as well as negative powers of $R$ (see Eq.(\ref{combined powers})). Due to co-existence of two different mass scales in (\ref{combined powers}), for an appropriated region in the parameter space, these models could provide a nice scenario where to address in a unified picture both, early-time inflation, and late-time speed-up.

The authors aknowledge the MES of Cuba for partial support of the research. The work of I Q was partly supported by CONACyT M\'exico grant number I0101/131/07 C-234/07, Instituto Avanzado de Cosmologia (IAC) collaboration.

\end{document}